\documentclass[preprint,12pt,authoryear]{elsarticle}

\usepackage[utf8]{inputenc}
\usepackage[T1]{fontenc}
\usepackage[spanish]{babel}
\usepackage{booktabs}
\usepackage{newpxtext,newpxmath}
\usepackage{microtype}
\usepackage{xcolor} 
\usepackage{tcolorbox} 
\usepackage{setspace}
\usepackage{geometry}
\usepackage{url}
\usepackage{fancyhdr}   
\usepackage{titlesec}   
\usepackage{footmisc}   
\usepackage{array} 

\renewcommand{\thesection}{\Roman{section}}
\renewcommand{\thesubsection}{\thesection.\Alph{subsection}}
\renewcommand{\thesubsubsection}{\thesubsection.\arabic{subsubsection}}
\definecolor{sectioncolor}{RGB}{70,70,140}
\definecolor{subsectioncolor}{RGB}{80,120,180}
\definecolor{subsubsectioncolor}{RGB}{100,140,200}

\newcommand{\sectionline}{\noindent\rule{\textwidth}{0.5pt}}

\titleformat{\section}{\normalfont\Large\bfseries\color{sectioncolor}}{\thesection}{1em}{}
\titleformat{\subsection}{\normalfont\large\bfseries\color{subsectioncolor}}{\thesubsection}{1em}{}
\titleformat{\subsubsection}{\normalfont\normalsize\bfseries\color{subsubsectioncolor}}{\thesubsubsection}{1em}{}

\titlespacing*{\section}{0pt}{2ex plus 1ex minus .2ex}{1ex}
\titlespacing*{\subsection}{0pt}{2ex plus 0.5ex minus .2ex}{0.5ex}
\titlespacing*{\subsubsection}{0pt}{1.5ex plus 0.2ex minus .2ex}{0.3ex}

\newcommand{\Section}[1]{\section{#1}\sectionline}

\biboptions{authoryear}

\begin{document}
	
	\begin{center}
		{\LARGE \textbf{La ley del descenso tendencial de la tasa de ganancia: Evidencia empírica para la economía española}}\\[0.5em]
		\sectionline
	\end{center}
	
	\begin{tcolorbox}[colback=gray!5!white,colframe=gray!40!black,
		boxrule=0.5pt,arc=3pt, left=5pt,right=5pt,top=5pt,bottom=5pt]
		\textbf{Iván López-Espejo}\footnote{Autor de correspondencia: iloes@ugr.es}\\
		Departamento de Teoría de la Señal, Telemática y Comunicaciones,\\
		Universidad de Granada, 18071 Granada, España
	\end{tcolorbox}
	
	\begin{tcolorbox}[colback=gray!5!white,colframe=gray!40!black,
		boxrule=0.5pt,arc=3pt, left=5pt,right=5pt,top=5pt,bottom=5pt]
		\textbf{Resumen:} \\
		Este artículo examina la ley del descenso tendencial de la tasa de ganancia en la economía española entre 1960 y 2024, considerando la composición orgánica del capital y la tasa de plusvalía como variables centrales. Su objetivo es determinar si esta ley, formulada por Marx en <<El Capital>> (Vol. III), sigue operando en el contexto contemporáneo. La metodología consiste en transformar categorías macroeconómicas ortodoxas derivadas de la Contabilidad Nacional de España (CNE), disponibles en BDMACRO, en variables marxistas: capital constante ($c$), capital variable ($v$) y plusvalía ($pv$). A partir de estas, se construyen series históricas de la composición orgánica del capital ($q$), la tasa de plusvalía ($pv'$) y la tasa de ganancia ($g'$), ajustadas a precios constantes para garantizar coherencia y comparabilidad temporal. Los resultados muestran un aumento sostenido de $q$ y una ligera disminución de $pv'$, generando un descenso tendencial de $g'$, con fluctuaciones cíclicas asociadas a crisis puntuales. Las conclusiones confirman empíricamente la validez de la ley en España, constatando los límites históricos del capitalismo y aportando evidencia cuantitativa sobre la dinámica estructural de la rentabilidad.
		
		\textbf{Palabras clave:} Tendencia de la tasa de ganancia, composición orgánica del capital, tasa de plusvalía, economía española, ciclos económicos
	\end{tcolorbox}
	
	\begin{tcolorbox}[colback=gray!5!white,colframe=gray!40!black,
		boxrule=0.5pt,arc=3pt, left=5pt,right=5pt,top=5pt,bottom=5pt]
		\textbf{Abstract:} \\
		This article examines the law of the tendency of the rate of profit to fall in the Spanish economy between 1960 and 2024, considering the organic composition of capital and the rate of surplus value as central variables. Its aim is to determine whether this law, formulated by Marx in \emph{Capital} (Vol. III), continues to operate in the contemporary context. The methodology consists of transforming orthodox macroeconomic categories derived from the Spanish National Accounts (CNE), available in BDMACRO, into Marxist variables: constant capital ($c$), variable capital ($v$), and surplus value ($pv$). Based on these, historical series of the organic composition of capital ($q$), the rate of surplus value ($pv'$), and the rate of profit ($g'$) are constructed, adjusted to constant prices to ensure temporal coherence and comparability. The results show a sustained increase in $q$ and a slight decrease in $pv'$, generating a tendential decline in $g'$ with cyclical fluctuations associated with specific crises. The conclusions empirically confirm the validity of the law in Spain, highlighting the historical limits of capitalism and providing quantitative evidence on the structural dynamics of profitability.
		
	\end{tcolorbox}
	
	\Section{Introducción}
	\label{sec:intro}
	El modo de producción capitalista, actualmente dominante a escala global, presenta límites históricos vinculados al alto grado de desarrollo de las fuerzas productivas que este sistema genera \citep{Harvey1999}. Formulada por Karl Marx y publicada en el Volumen III de <<El Capital>> \citep{ElCapitalIII}, la ley del descenso tendencial de la tasa de ganancia constituye una de las expresiones más acabadas de estos límites históricos. En este breve artículo, llevamos a cabo un análisis empírico de la evolución de la tasa de ganancia y sus categorías económicas integrantes (es decir, la composición orgánica del capital y la tasa de plusvalía) en el contexto de la economía española, con el objetivo primordial de determinar si la ley del descenso tendencial de la tasa de ganancia sigue operando en este marco. Cabe señalar que, si bien \cite{Izquierdo03} realizó un estudio similar al aquí propuesto, una de las diferencias más importantes radica en que en este trabajo se contempla un período de análisis mucho más amplio (1960--2024 vs. 1964--2000). Esto es crucial debido a la naturaleza tendencial de la ley que nos ocupa, pudiendo fenómenos coyunturales distorsionar su visibilidad en el corto plazo \citep{ElCapitalIII,Harvey2012}.
	
	\subsection{Definición de las categorías esenciales}
	En primera instancia, recordemos que es posible considerar el conjunto de la economía de una determinada sociedad capitalista como la de una gran empresa, \emph{metafóricamente hablando} \citep{ElCapitalIII,Carchedi2011,Kliman2011}. Por tanto, la tasa de ganancia o rentabilidad de dicha economía, $g'$, viene dada por la expresión
	\begin{equation}
		g'=\displaystyle\frac{pv}{c+v},
		\label{eq:tasagan}
	\end{equation}
	donde $pv$ es la plusvalía (la cual podemos aproximar en nuestro contexto al total de beneficios generados en la economía bajo análisis), $c$ es el capital constante (capital total invertido en medios de producción) y $v$ es el capital variable (capital total invertido en fuerza de trabajo).
	
	Seguidamente, estamos en disposición de definir la composición orgánica del capital (que puede interpretarse como un indicador del grado de tecnificación de la producción en un contexto determinado), $q$, como el cociente entre el capital constante, $c$, y el capital variable, $v$, es decir,
	\begin{equation}
		q=\displaystyle\frac{c}{v}.
		\label{eq:coc}
	\end{equation}
	
	De otra parte, definimos la tasa de plusvalía (la cual refleja la intensidad de explotación del trabajo asalariado), $pv'$, como la relación entre la plusvalía, $pv$, y el capital invertido en fuerza de trabajo, $v$:
	\begin{equation}
		pv'=\displaystyle\frac{pv}{v}.
		\label{eq:tasaplus}
	\end{equation}
	
	Finalmente, nótese que podemos reescribir la expresión de la tasa de ganancia o rentabilidad recogida en la Ecuación (\ref{eq:tasagan}) en términos de la tasa de plusvalía, $pv'$, y de la composición orgánica del capital, $q$, como
	\begin{equation}
		g'=\displaystyle\frac{pv}{c+v}=\frac{pv'}{1+q}.
		\label{eq:tasagan2}
	\end{equation}
	
	En la subsección siguiente nos apoyamos en las categorías económicas marxistas, derivadas de la teoría del valor-trabajo, aquí definidas para ilustrar la ley del descenso tendencial de la tasa de ganancia.
	
	\subsection{Ley del descenso tendencial de la tasa de ganancia}
	\label{ssec:ldttg}
	Como ya se teorizara, la composición orgánica del capital, $q$, tiende a aumentar con el transcurso del tiempo en el marco del modo de producción capitalista \citep{ElCapitalIII,Kliman2011,Carchedi2011}. La explicación a este fenómeno es la siguiente. Con el objetivo de mejorar su posición competitiva en el mercado, cada capitalista, \emph{de forma individual}, busca incrementar la productividad a través de invertir en la automatización o mecanización de sus procesos de producción. Sin embargo, \emph{en términos agregados} (es decir, cuando el conjunto de la clase capitalista se mueve en la dirección de la automatización como resultado de la competencia), esta dinámica conlleva un aumento de la inversión social total en medios de producción ($c\!\uparrow$), así como una posible reducción de la fuerza de trabajo social total ($v\!\downarrow$). En otras palabras, la mecanización de los procesos productivos derivada de la pugna intercapitalista provoca el crecimiento, \emph{como tendencia}, de la composición orgánica del capital ($q\!\uparrow$, véase la Ecuación (\ref{eq:coc})) a lo largo del tiempo \citep{ElCapitalIII,Kliman2011,Carchedi2011}.
	
	Con base en la Ecuación (\ref{eq:tasagan2}), que expresa $g'$ en función de $q$, resulta evidente que la evolución creciente de la composición orgánica del capital presiona a la baja la tasa de ganancia ($g'\!\downarrow$), lo que sólo puede compensarse \emph{de manera limitada y temporal} mediante el incremento de la explotación del trabajo asalariado (es decir, aumentando $pv'$). Este fenómeno es lo que se conoce como ley del descenso tendencial de la tasa de ganancia, ley que actúa como uno de los mecanismos estructurales más relevantes en la génesis de crisis económicas (en tanto que ralentización o interrupción del ritmo de acumulación) de escasez de plusvalía \citep{ElCapitalIII,Kliman2011,Carchedi2011,Roberts2016}.
	
	Como corolario, obsérvese el carácter contradictorio del comportamiento aquí descrito, que es a su vez causa de la progresión de $q$: la automatización de los procesos productivos en la que cada capitalista invierte individualmente para mejorar su posición competitiva en el mercado y, por tanto, su ganancia, en términos agregados hace reducir, en su propio perjuicio, la rentabilidad que obtiene el conjunto de los capitalistas como clase social.
	
	Esta contradicción fundamental será examinada en la evolución histórica de la economía española en las siguientes secciones.
	
	\subsection{Resumen de la organización del artículo}
	El resto del artículo se organiza de la siguiente manera: la Sección \ref{sec:metod} describe la metodología seguida para elaborar curvas de composición orgánica del capital y de tasas de plusvalía y ganancia a partir de categorías económicas ortodoxas derivadas de la Contabilidad Nacional de España (CNE), tal y como se presentan en la base de datos BDMACRO \citep{BDMACRO2025}. Las curvas obtenidas se muestran y discuten en la Sección \ref{sec:resultados}. Finalmente, la Sección \ref{sec:conclusiones} concluye el artículo.
	
	\Section{Metodología}
	\label{sec:metod}
	Para construir curvas de composición orgánica del capital y de tasas de plusvalía y ganancia, recurrimos en este trabajo a BDMACRO \citep{BDMACRO2025}, una base de datos anual de magnitudes macroeconómicas de la economía española. Estas series proceden de la CNE y tienen año base 2020. La razón primordial para la elección de BDMACRO radica en el amplio período temporal que abarca (1954--2024) en comparación con otras bases de datos de magnitudes macroeconómicas españolas.
	
	El primer paso consiste en establecer una correspondencia entre las categorías económicas ortodoxas de BDMACRO y las categorías marxistas elementales $c$, $v$ y $pv$ que posibilitan el cálculo de $q$, $pv'$ y $g'$ (véanse las Ecuaciones (\ref{eq:coc}), (\ref{eq:tasaplus}) y (\ref{eq:tasagan2})):
	\begin{enumerate}
		\item \emph{Capital constante, $c$}: Empleamos como medida de esta variable el \emph{stock} de capital fijo \citep{Kliman2011,Carchedi2011,Roberts2016} dado por la categoría de BDMACRO \texttt{Stock de Capital total (AMECO)}. Cabe señalar que esta serie (1960--2024) proviene de la base de datos AMECO de la Comisión Europea \citep{EC2025}.
		\item \emph{Capital variable, $v$}: El valor de la fuerza de trabajo es aproximado aquí a la remuneración de asalariados.
		\item \emph{Plusvalía, $pv$}: Para la estimación de esta variable, seguimos una de las metodologías habitualmente utilizadas en la literatura marxista contemporánea consistente en sustraerle al producto interior bruto (PIB) el capital variable, $v$, y el consumo de capital fijo (es decir, el total de los activos fijos consumidos como consecuencia del desgaste natural a lo largo del período en cuestión), $c'$ \citep{Kliman2011,Carchedi2011,Roberts2016}. En términos matemáticos,
		\begin{equation}
			pv\approx \mbox{PIB}-v-c'.
			\label{eq:pv_aprox}
		\end{equation}
	\end{enumerate}
	
	Es esencial aclarar que los diferentes flujos anuales de BDMACRO aquí considerados (es decir, el PIB, la remuneración de asalariados y el consumo de capital fijo) están expresados en precios corrientes. Antes de emplearlos, estos flujos se convierten a precios constantes haciendo uso del deflactor\footnote{Un deflactor es un índice económico que se utiliza para eliminar el efecto de la inflación en el valor monetario de los bienes y servicios.} del PIB con año base 2020 proporcionado por la misma base de datos, ya que constituye el indicador más adecuado entre los disponibles. La aplicación de este procedimiento de deflactación es clave para, por un lado, garantizar la coherencia entre las distintas categorías económicas ortodoxas empleadas en este artículo, puesto que el \emph{stock} de capital fijo (es decir, \texttt{Stock de Capital total (AMECO)}) viene indicado en precios constantes. Por otro lado, el trabajar a precios constantes nos permite analizar tendencias históricas y ciclos sin que la inflación distorsione la interpretación.
	
	\begin{table}[tbp]
		\centering
		\caption{Resumen de las magnitudes macroeconómicas de BDMACRO consideradas en este artículo y su modo de uso para estimar las categorías económicas marxistas elementales $c$, $v$ y $pv$.}
		\begin{tabular}{p{5cm} p{3cm} p{7cm}}
			\toprule
			\textbf{Magnitud} & \textbf{Sigla/código} & \textbf{Modo de uso} \\
			\midrule
			\texttt{Stock de Capital total (AMECO)} & \textbf{K\_}$_{\mbox{\scriptsize \textbf{AMECO}}}$ &
			Uso como capital constante $c$ \\ \midrule
			
			\texttt{Remuneración de asalariados} & \textbf{RA} &
			Tras deflactar el dato, uso como capital variable $v$ \\ \midrule
			
			\texttt{Producto Interior Bruto a precios corrientes} & \textbf{PIBpm} &
			Tras deflactar el dato, uso en Ecuación (\ref{eq:pv_aprox}) para aproximar $pv$ \\ \midrule
			
			\texttt{Consumo de capital fijo} & \textbf{CCF} &
			Tras deflactar el dato, uso como $c'$ en Ecuación (\ref{eq:pv_aprox}) para aproximar $pv$ \\ \midrule
			
			\texttt{Deflactor del Producto Interior Bruto} & \textbf{dPIB} &
			Usado para transformar a precios constantes los flujos \textbf{RA}, \textbf{PIBpm} y \textbf{CCF} \\ 
			\bottomrule
		\end{tabular}
		\label{tab:bdmacro}
	\end{table}
	
	Nótese que todos los flujos anuales y el \emph{stock} de capital fijo procedentes de BDMACRO utilizados en este estudio se expresan en millones de euros y se usan directamente en dicha unidad monetaria, siguiendo la literatura previa, que sugiere que los precios pueden emplearse como una medida \emph{proxy} razonable de los tiempos de trabajo \citep{Izquierdo03,Zhao2024}.
	
	Finalmente, para mayor conveniencia del lector, el Cuadro \ref{tab:bdmacro} recoge un resumen de las magnitudes macroeconómicas de BDMACRO adoptadas en el presente trabajo y su modo de uso para aproximar las categorías económicas marxistas elementales $c$, $v$ y $pv$. A partir de estas categorías se construyen las curvas de composición orgánica del capital y de tasas de plusvalía y ganancia (Ecuaciones (\ref{eq:coc}), (\ref{eq:tasaplus}) y (\ref{eq:tasagan2}), respectivamente), las cuales se muestran en el apartado siguiente.
	
	\Section{Resultados y discusión}
	\label{sec:resultados}
	
	Esta sección se organiza en dos partes: la presentación de los resultados (Subsección \ref{ssec:resultados}) y su discusión e interpretación (Subsección \ref{ssec:discussion}), centradas en la evolución de la tasa de ganancia y de sus categorías económicas constitutivas en el contexto de la economía española.
	
	\subsection{Presentación de resultados}
	\label{ssec:resultados}
	
	\begin{figure}
		\includegraphics[width=\linewidth]{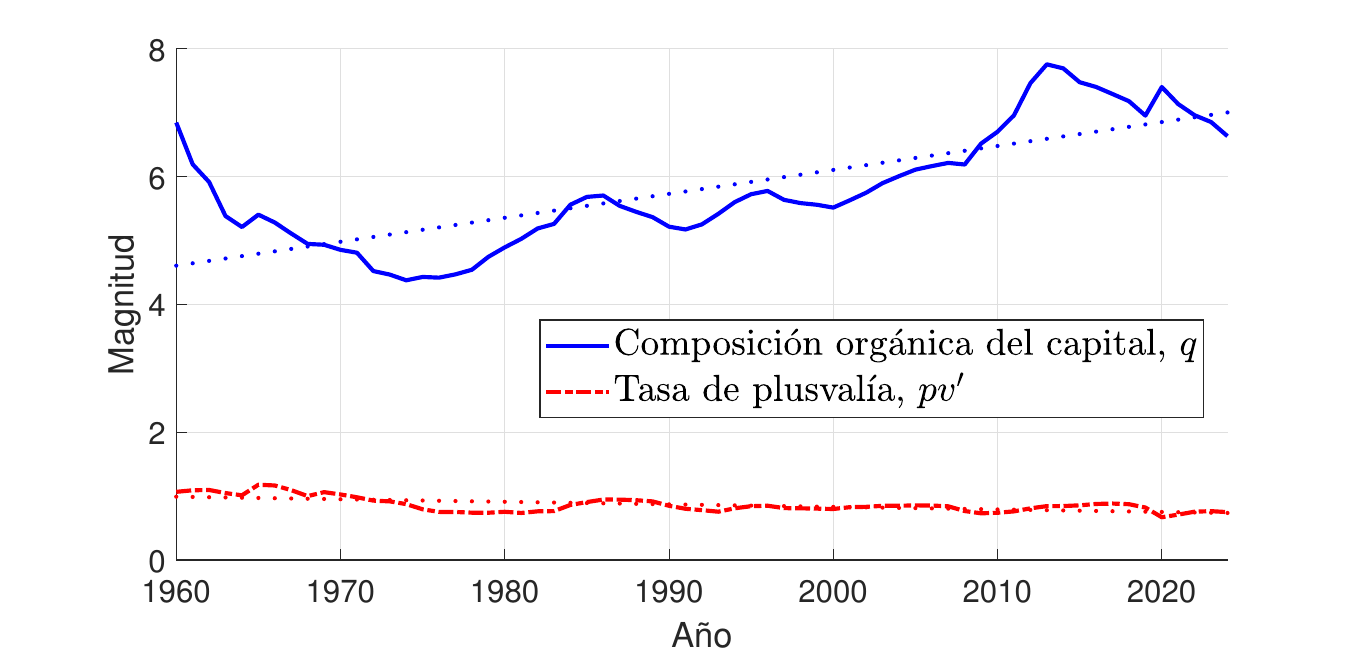}
		\caption{Evolución de la composición orgánica del capital, $q$, y de la tasa de plusvalía, $pv'$, para el conjunto de la economía española desde 1960 hasta 2024. Las tendencias de estas curvas (líneas punteadas) se han obtenido mediante sendos ajustes lineales por mínimos cuadrados. Fuente: elaboración propia a partir de BDMACRO.}
		\label{fig:COCyTP}
	\end{figure}
	
	\begin{figure}
		\includegraphics[width=\linewidth]{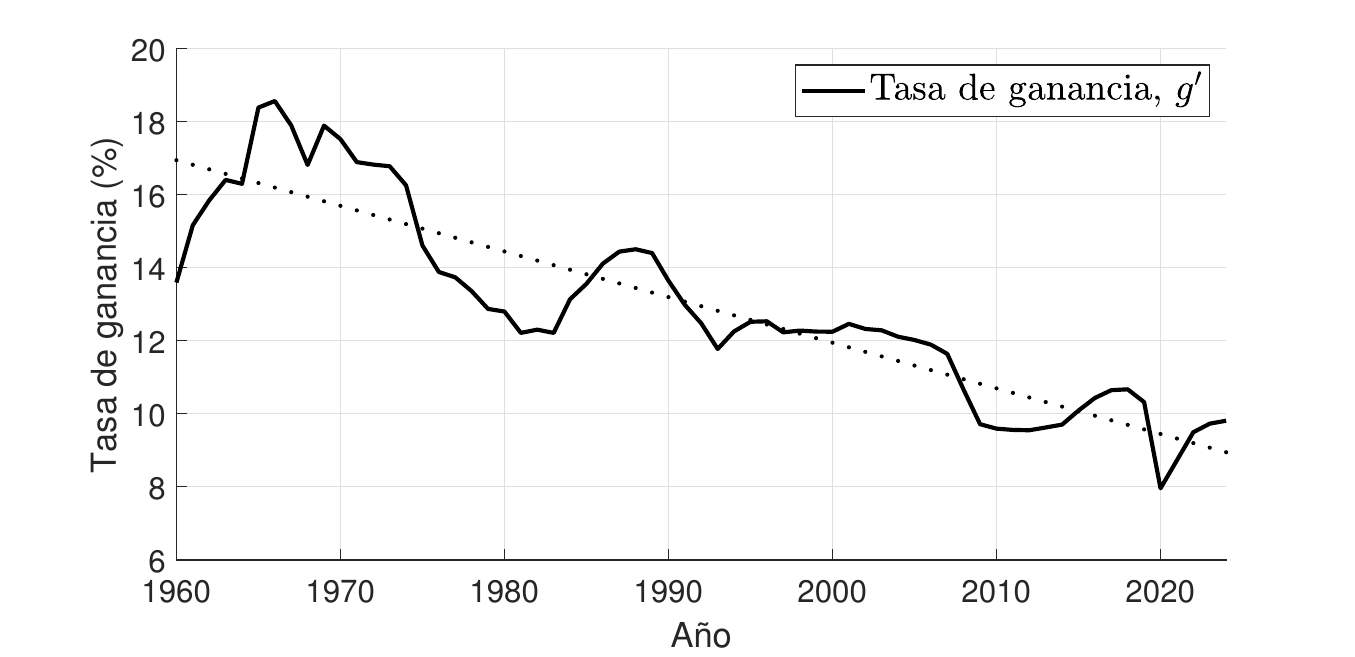}
		\caption{Evolución de la tasa de ganancia, $g'$, en porcentaje, para el conjunto de la economía española desde 1960 hasta 2024. La tendencia de esta curva (línea punteada) se ha obtenido mediante un ajuste lineal por mínimos cuadrados. Fuente: elaboración propia a partir de BDMACRO.}
		\label{fig:tasagan}
	\end{figure}
	
	La Figura \ref{fig:COCyTP} representa la evolución de la composición orgánica del capital, $q$, y de la tasa de plusvalía, $pv'$, en el ámbito de la economía española entre 1960 y 2024. De forma paralela, la Figura \ref{fig:tasagan} ilustra la dinámica temporal de la tasa de ganancia, $g'$, en España para el mismo intervalo de tiempo. Este marco temporal de análisis (es decir, 1960--2024) se seleccionó debido a que es el mayor lapso de tiempo para el cual existen datos disponibles en BDMACRO de todas las magnitudes macroeconómicas necesarias. Además, obsérvese que estas mismas figuras también representan tendencias lineales ---estimadas mediante mínimos cuadrados ordinarios (MCO) \citep{Wooldridge2013,Gujarati2009}--- de las diferentes categorías económicas ($q$, $pv'$ y $g'$).
	
	El Cuadro \ref{tab:resultados} recopila los datos numéricos (redondeados al segundo decimal) empleados en la generación de las anteriores figuras. Adicionalmente, con el objetivo de facilitar la reproducibilidad del presente estudio, el Cuadro \ref{tab:c_v_pv} recoge los valores numéricos anuales, en millones de euros, de las categorías marxistas elementales $c$, $v$ y $pv$ inicialmente estimadas.
	
	Todos los resultados expuestos en esta subsección se han obtenido siguiendo la metodología detallada en la Sección \ref{sec:metod}.
	
	\begin{table}[tbp]
		\centering
		\caption{Valores anuales de la composición orgánica del capital ($q$), la tasa de plusvalía ($pv'$) y la tasa de ganancia ($g'$), esta última en porcentaje, para el conjunto de la economía española. Todos los valores han sido redondeados al segundo decimal. Fuente: elaboración propia a partir de BDMACRO.}
		\begin{tabular}{c c c c | c c c c}
			\toprule
			\textbf{Año} & \textbf{$q$} & \textbf{$pv'$} & \textbf{$g'$ (\%)} &
			\textbf{Año} & \textbf{$q$} & \textbf{$pv'$} & \textbf{$g'$ (\%)} \\
			\midrule
			1960 & 6,84 & 1,07 & 13,60 & 1993 & 5,42 & 0,76 & 11,78 \\ 
			1961 & 6,19 & 1,09 & 15,17 & 1994 & 5,60 & 0,81 & 12,26 \\ 
			1962 & 5,91 & 1,10 & 15,85 & 1995 & 5,72 & 0,84 & 12,52 \\ 
			1963 & 5,38 & 1,05 & 16,41 & 1996 & 5,77 & 0,85 & 12,54 \\ 
			1964 & 5,21 & 1,01 & 16,30 & 1997 & 5,64 & 0,81 & 12,23 \\ 
			1965 & 5,40 & 1,18 & 18,39 & 1998 & 5,58 & 0,81 & 12,28 \\ 
			1966 & 5,28 & 1,17 & 18,57 & 1999 & 5,56 & 0,80 & 12,26 \\ 
			1967 & 5,11 & 1,09 & 17,90 & 2000 & 5,51 & 0,80 & 12,25 \\ 
			1968 & 4,95 & 1,00 & 16,82 & 2001 & 5,63 & 0,83 & 12,47 \\ 
			1969 & 4,93 & 1,06 & 17,89 & 2002 & 5,75 & 0,83 & 12,33 \\ 
			1970 & 4,85 & 1,03 & 17,53 & 2003 & 5,90 & 0,85 & 12,29 \\ 
			1971 & 4,81 & 0,98 & 16,90 & 2004 & 6,01 & 0,85 & 12,11 \\ 
			1972 & 4,52 & 0,93 & 16,83 & 2005 & 6,11 & 0,85 & 12,02 \\ 
			1973 & 4,47 & 0,92 & 16,78 & 2006 & 6,16 & 0,85 & 11,90 \\ 
			1974 & 4,38 & 0,87 & 16,26 & 2007 & 6,21 & 0,84 & 11,64 \\ 
			1975 & 4,43 & 0,79 & 14,61 & 2008 & 6,19 & 0,77 & 10,66 \\ 
			1976 & 4,42 & 0,75 & 13,89 & 2009 & 6,52 & 0,73 & 9,72 \\ 
			1977 & 4,47 & 0,75 & 13,74 & 2010 & 6,70 & 0,74 & 9,60 \\ 
			1978 & 4,54 & 0,74 & 13,36 & 2011 & 6,96 & 0,76 & 9,56 \\ 
			1979 & 4,74 & 0,74 & 12,88 & 2012 & 7,46 & 0,81 & 9,55 \\ 
			1980 & 4,89 & 0,75 & 12,80 & 2013 & 7,76 & 0,84 & 9,63 \\ 
			1981 & 5,02 & 0,74 & 12,22 & 2014 & 7,69 & 0,84 & 9,71 \\ 
			1982 & 5,19 & 0,76 & 12,31 & 2015 & 7,48 & 0,86 & 10,09 \\ 
			1983 & 5,26 & 0,76 & 12,22 & 2016 & 7,40 & 0,88 & 10,44 \\ 
			1984 & 5,56 & 0,86 & 13,14 & 2017 & 7,29 & 0,88 & 10,65 \\ 
			1985 & 5,68 & 0,91 & 13,56 & 2018 & 7,18 & 0,87 & 10,67 \\ 
			1986 & 5,70 & 0,95 & 14,12 & 2019 & 6,95 & 0,82 & 10,32 \\ 
			1987 & 5,54 & 0,94 & 14,44 & 2020 & 7,40 & 0,67 & 7,96 \\ 
			1988 & 5,45 & 0,94 & 14,51 & 2021 & 7,13 & 0,71 & 8,73 \\ 
			1989 & 5,36 & 0,92 & 14,41 & 2022 & 6,96 & 0,76 & 9,50 \\ 
			1990 & 5,22 & 0,85 & 13,65 & 2023 & 6,85 & 0,76 & 9,73\\ 
			1991 & 5,17 & 0,80 & 12,99 & 2024 & 6,63 & 0,75 & 9,81 \\ 
			1992 & 5,25 & 0,78 & 12,48 & --- & --- & --- & --- \\
			\bottomrule
		\end{tabular}
		\label{tab:resultados}
	\end{table}
	
	\begin{table}[tbp]
		\centering
		\caption{Valores anuales de capital constante ($c$), capital variable ($v$) y plusvalía ($pv$), expresados en millones de euros, para el conjunto de la economía española. Los valores de $v$ y de $pv$ han sido redondeados a la unidad. Fuente: elaboración propia a partir de BDMACRO.}
		\begin{tabular}{c c c c | c c c c}
			\toprule
			\textbf{Año} & \textbf{$c$} & \textbf{$v$} & \textbf{$pv$} &
			\textbf{Año} & \textbf{$c$} & \textbf{$v$} & \textbf{$pv$} \\
			\midrule
			1960 & 524.949 & 76.696 & 81.804 & 1993 & 1.967.049 & 363.175 & 274.475 \\ 
			1961 & 536.655 & 86.672 & 94.548 & 1994 & 2.022.678 & 361.187 & 292.234 \\ 
			1962 & 551.122 & 93.178 & 102.107 & 1995 & 2.086.446 & 364.496 & 306.945 \\ 
			1963 & 568.731 & 105.679 & 110.659 & 1996 & 2.150.686 & 372.428 & 316.309 \\ 
			1964 & 589.973 & 113.239 & 114.632 & 1997 & 2.220.024 & 393.909 & 319.770 \\ 
			1965 & 616.983 & 114.179 & 134.448 & 1998 & 2.302.539 & 412.279 & 333.438 \\ 
			1966 & 648.918 & 122.886 & 143.308 & 1999 & 2.398.522 & 431.483 & 346.822 \\ 
			1967 & 684.163 & 133.913 & 146.436 & 2000 & 2.504.843 & 454.257 & 362.520 \\ 
			1968 & 721.907 & 145.960 & 145.952 & 2001 & 2.615.250 & 464.684 & 383.956 \\ 
			1969 & 764.242 & 154.930 & 164.482 & 2002 & 2.729.984 & 475.034 & 395.087 \\ 
			1970 & 806.908 & 166.267 & 170.552 & 2003 & 2.854.796 & 484.090 & 410.321 \\ 
			1971 & 845.710 & 175.876 & 172.606 & 2004 & 2.985.777 & 497.152 & 421.938 \\ 
			1972 & 893.723 & 197.669 & 183.657 & 2005 & 3.129.202 & 512.288 & 437.859 \\ 
			1973 & 951.124 & 212.966 & 195.366 & 2006 & 3.286.964 & 533.286 & 454.498 \\ 
			1974 & 1.012.912 & 231.430 & 202.334 & 2007 & 3.449.818 & 555.188 & 466.356 \\ 
			1975 & 1.068.347 & 241.210 & 191.357 & 2008 & 3.593.358 & 580.645 & 444.876 \\ 
			1976 & 1.121.085 & 253.746 & 190.927 & 2009 & 3.681.713 & 564.987 & 412.721 \\ 
			1977 & 1.171.793 & 262.158 & 197.035 & 2010 & 3.754.000 & 560.185 & 414.039 \\ 
			1978 & 1.219.172 & 268.439 & 198.777 & 2011 & 3.805.411 & 546.914 & 416.173 \\ 
			1979 & 1.260.500 & 265.739 & 196.517 & 2012 & 3.837.356 & 514.294 & 415.721 \\ 
			1980 & 1.301.701 & 266.077 & 200.730 & 2013 & 3.858.774 & 497.516 & 419.433 \\ 
			1981 & 1.339.028 & 266.499 & 196.183 & 2014 & 3.885.605 & 505.091 & 426.232 \\ 
			1982 & 1.375.501 & 265.150 & 201.923 & 2015 & 3.919.638 & 524.282 & 448.519 \\ 
			1983 & 1.408.728 & 267.851 & 204.860 & 2016 & 3.954.277 & 534.201 & 468.473 \\ 
			1984 & 1.434.893 & 258.070 & 222.472 & 2017 & 3.999.063 & 548.432 & 484.319 \\ 
			1985 & 1.464.373 & 257.735 & 233.574 & 2018 & 4.054.816 & 564.768 & 493.082 \\ 
			1986 & 1.502.300 & 263.402 & 249.233 & 2019 & 4.119.219 & 592.349 & 486.402 \\ 
			1987 & 1.550.366 & 279.797 & 264.350 & 2020 & 4.158.324 & 561.866 & 375.852 \\ 
			1988 & 1.610.714 & 295.651 & 276.599 & 2021 & 4.201.219 & 588.882 & 418.169 \\ 
			1989 & 1.683.582 & 313.819 & 287.727 & 2022 & 4.249.952 & 610.672 & 461.715 \\ 
			1990 & 1.762.541 & 337.928 & 286.677 & 2023 & 4.298.101 & 627.175 & 479.405 \\ 
			1991 & 1.841.478 & 356.050 & 285.377 & 2024 & 4.348.757 & 655.728 & 491.123 \\ 
			1992 & 1.911.864 & 364.046 & 284.034 & --- & --- & --- & --- \\ 
			\bottomrule
		\end{tabular}
		\label{tab:c_v_pv}
	\end{table}
	
	\subsection{Discusión}
	\label{ssec:discussion}
	
	Tal y como se observa en la Figura \ref{fig:COCyTP}, la composición orgánica del capital muestra una tendencia claramente ascendente como resultado de la creciente mecanización de los procesos productivos señalada en la Subsección \ref{ssec:ldttg}. Sin embargo, la tasa de plusvalía ---cuya determinación se inscribe en el plano de la lucha de clases \citep{Vol1Capital,Mandel1979}--- experimenta un ligero descenso a lo largo del tiempo a pesar de la evidente degradación de las condiciones materiales de la clase trabajadora española a través de estas últimas décadas \citep{Ubeda2020}. La explicación de este fenómeno \emph{aparentemente} contradictorio radica en que, en economías como la española, donde el peso del sector inmobiliario y financiero es muy elevado, parte de la extracción de valor se desplaza de la esfera productiva hacia mecanismos financieros y rentistas, lo cual no tiene un reflejo directo en la tasa clásica de plusvalía \citep{Lapavitsas2013,LopezRodriguez2011}. Entre estos mecanismos que transfieren rentas del trabajo al capital sin necesariamente pasar por la plusvalía productiva clásica se encuentran el aumento del endeudamiento de los hogares (p. ej., hipotecas y créditos al consumo), inflación en vivienda y alquileres (que actualmente absorben una proporción considerable del salario \citep{Romero2024}) o la privatización de servicios públicos como sanidad y educación. En consecuencia, resulta necesario considerar de manera integrada las esferas financiero-rentista y productiva para, empíricamente (mediante categorías distintas de $pv'$), estimar el aumento real de la explotación de la clase trabajadora \citep{Farina2018}.
	
	La evolución creciente de $q$, combinada con el comportamiento más o menos estable de $pv'$, revela, a partir de la Ecuación (\ref{eq:tasagan2}), la tendencia descendente de la tasa de ganancia representada en la Figura \ref{fig:tasagan}. Esta tendencia de la rentabilidad a disminuir (14--18\% al comienzo de la serie vs. 8--10\% al final de la misma) con la consecución de los ciclos productivos de la economía española como consecuencia de la trayectoria ascendente de $q$ respalda la validez de la ley del descenso tendencial de la tasa de ganancia \citep{ElCapitalIII}. Nótese que este declive estructural de $g'$ es la causa subyacente del desplazamiento de la extracción de valor desde la esfera productiva (progresivamente menos rentable) hacia la esfera financiero-rentista \citep{Tome2021}, tal como se discutió en el párrafo anterior.
	
	Más allá de la tendencia descendente de la curva de la tasa de ganancia, resulta relevante atender a su contorno, ya que, a través de sus fluctuaciones cíclicas de corto plazo, refleja una diversidad de acontecimientos históricos. En particular, los valles de $g'$ en la Figura \ref{fig:tasagan} constituyen indicadores de etapas de depresión de la economía española. Entre ellos, cabe destacar: \emph{1)} el agotamiento del crecimiento de posguerra y el auge de las políticas neoliberales como respuesta (1973--1983), \emph{2)} la crisis económica en España de 1992--1993 \citep{Fuentes1993}, \emph{3)} la Gran Recesión (2008--2013), y \emph{4)} el impacto de la pandemia de COVID-19 (2020).
	
	Un análisis comparativo de las Figuras \ref{fig:COCyTP} y \ref{fig:tasagan} revela asimismo las siguientes correlaciones:
	\begin{enumerate}
		\item \emph{Entre la tasa de plusvalía y la tasa de ganancia}: Coincidiendo con las fases de crisis económica (valles de $g'$), la tasa de plusvalía, $pv'$, tiende a experimentar ligeros descensos. Este fenómeno no se debe a que la clase capitalista relaje la presión sobre la fuerza de trabajo, sino a que la plusvalía efectiva ---que se estima a través del PIB; véase la Ecuación (\ref{eq:pv_aprox})--- disminuye temporalmente como consecuencia de las dificultades en su realización en el mercado y de la crisis general \citep{Shaikh2016,Kliman2011}.
		\item \emph{Entre la composición orgánica del capital y la tasa de ganancia}: Coincidiendo con las fases de desaceleración (inicio de los valles de $g'$), la composición orgánica del capital, $q$, tiende a experimentar un relanzamiento cuya magnitud varía según el contexto económico. Este comportamiento se explica principalmente porque la caída rápida del capital variable, $v$ ---a través de despidos y precarización--- eleva la proporción de capital constante, $c$, en relación al total invertido, sin que necesariamente haya un aumento real de la inversión en maquinaria \citep{Kliman2011,Roberts2016}.
	\end{enumerate}
	Estos descensos transitorios de $pv'$, combinados con la tendencia al alza de la composición orgánica del capital, explican en parte los valles observados en la serie de la tasa de ganancia, reforzando la validez empírica de la ley del descenso tendencial de la tasa de ganancia formulada por Marx \citep{Kliman2011,Roberts2016}.
	
	\Section{Conclusión}
	\label{sec:conclusiones}
	
	En conclusión, el presente estudio ha permitido validar empíricamente la ley del descenso tendencial de la tasa de ganancia en el caso de la economía española a lo largo del período 1960--2024. La evolución observada de la composición orgánica del capital y de la tasa de plusvalía, así como su relación con la tasa de ganancia, confirma que el crecimiento sostenido de $q$ ha ejercido una presión estructural sobre $g'$, mientras que la ligera tendencia descendente de $pv'$ refleja, de forma secundaria, la transferencia de extracción de valor desde la esfera productiva hacia mecanismos de carácter financiero-rentista. Estos resultados aportan evidencia de los límites históricos del capitalismo en España, mostrando cómo la lógica de acumulación y mecanización del capital genera restricciones sistémicas sobre la rentabilidad agregada. Asimismo, la metodología seguida, basada en la construcción de series de $c$, $v$ y $pv$ a partir de las magnitudes macroeconómicas de BDMACRO y su transformación a precios constantes, proporciona una base rigurosa y reproducible para el análisis, consolidando la utilidad de esta aproximación para estudios comparativos y posteriores investigaciones sobre la dinámica de la rentabilidad en distintas economías capitalistas.
	
	\bibliographystyle{elsarticle-harv}
	\bibliography{Bibliog}
	
\end{document}